\documentclass[12pt,a4paper,english]{article}
\usepackage[english]{babel}

\begin{document}

\Large
\centerline{A note on the forced Burgers equation}

\vspace{1cm}

\normalsize
\vspace{2cm}
\centerline{S. Eule\footnote{yeahhart@uni-muenster.de, Tel.:+49-251-8334924, Fax: +49-251-8336328}, R. Friedrich}
\centerline{Institute of Theoretical Physics}
\centerline{Westf\"alische Wilhelms Universit\"at}
\centerline{M\"unster}
\centerline{Wilhelm-Klemm-Str. 9}
\centerline{G-48149 M\"unster}

\vspace{1cm}

\normalsize

\centerline{Abstract}

We obtain the exact solution
for the Burgers equation with a time dependent
forcing, which depends linearly on the 
spatial coordinate. For the case of a 
stochastic time dependence an exact expression for 
the joint probability distribution for the velocity fields at multiple
spatial points is obtained. A connection with stretched vortices in
hydrodynamic flows is discussed.\\  

PACS number(s): 02.50.-r, 05.40.-a, 05.90.-a, 47.32.Cc\\

Burgers equation \cite{Burgers}, \cite{Frisch}, \cite{Woy} 
has extensively been investigated 
as a model equation for the study of the problem of turbulence.
The homogeneous equation can be exactly solved by a Hopf-Cole transformation
\cite{Hopf}, \cite{Cole}
such that many properties of the spatio-temporal behaviour can be
obtained by analytical means. 
In recent years, the stochastically 
forced Burgers equation has attracted much interest due to the work
of Polyakov \cite{Polyakov}. He considered a white noise forcing
$f(x,t)$ with spatial correlations
$<f(x,t)f(x',t)>=\kappa(x-x')\delta(t-t')$. 
This equation has been found to exhibit
scaling behaviour. Polyakov suggested to use 
an operator product expansion in order to assess the scaling behaviour of 
the velocity increments.  

The purpose of the present note is to present
the general solution of the forced Burgers equation
\begin{eqnarray}\label{evol}
[\frac{\partial}{\partial t}+v(r,t) \frac{\partial}{ \partial r}]
v(r,t)
&=& \nu
\frac{\partial^2}{\partial r^2}v(r,t)+G(t) r \qquad ,
\end{eqnarray}
where $G(t)$ is an arbitrary function of time t. As we shall 
discuss below, this problem is related to the Polyakov problem. The 
investigation of the special type of forcing, $f(r,t)=rG(t)$, has the 
advantage that one can find an analytic solution for an arbitrary
function $G(t)$.

This solution can be found by direct calculation using 
the ansatz
\begin{equation}\label{ansatz}
v(r,t)=a(t) r + \Lambda(t) w(\Lambda(t) r,\tau(t)) \qquad . 
\end{equation}
Thereby, the functions $a(t)$, $\Lambda(t)$, $\tau(t)$ are determined by 
the following solvability conditions:
\begin{eqnarray}\label{diff}
\dot a(t)+a(t)^2 &=& G(t) \qquad 
\nonumber \\
\dot \Lambda &=&-a(t) \Lambda \qquad ,
\nonumber \\
\dot \tau &=& \Lambda^2 \qquad .
\end{eqnarray}
The quantities $\Lambda(t)$ and $\tau(t)$ can be determined as a function
of $a(t)$,
\begin{equation}
\Lambda(t)= e^{-\int_0^t dt' a(t')} \qquad ,
\qquad 
\tau = \int_0^t dt'  e^{-2\int_0^{t'} dt'' a(t'') }
\end{equation}
In the following we shall see that the quantity 
\begin{equation}
\frac{\Lambda^2(t)}{\tau(t)}=1/\sigma^2(t)=\frac{d}{dt} ln \tau(t)
\end{equation}
will play a major role.

In order that the ansatz eq. (\ref{ansatz}) obeys the forced Burgers equation
the function $w(\xi,\tau)$ has to obey the homogeneous equation. For instance,
$w(\xi,\tau)$ could be the single shock solution
\begin{equation}
w(\xi,\tau)=- A ~ tanh ~\frac{A}{2\nu} \xi
\end{equation}
Of interest are also scaling solutions of the form \cite{Woy}
\begin{equation}
w(\xi,\tau)=\frac{1}{\sqrt{\tau}} W(\frac{\xi}{\sqrt{\tau}}) \qquad,
\end{equation}
which e.g. evolve from the initial condition $w(\xi,0)=\delta(\xi)$. 

Since the solution of the homogeneous Burgers equation is known in terms
of the initial conditions 
the function $\Lambda(t) w(\xi,\tau)$ can be represented according to
\begin{equation}
\Lambda(t) w(\Lambda r,\tau)=-2 \nu \Lambda(t) \frac{\partial }{\partial r}
ln \int_{-\infty}^{\infty} dr'
e^{-(r-r')^2/(4\nu \sigma^2(t)) }e^{-\int_{-\infty}^{r'} dr' \Lambda(t) 
w(\Lambda r',0)/2\nu}
\end{equation}

We mention that the structure of the solution, 
eq. (\ref{ansatz}), is similar to the solution describing
stretched vortices in the Navier-Stokes equation (see e.g. \cite{Lundgren}).
The analogy becomes even more clear since, 
as we shall briefly mention below, the treatment 
can be extended to the case of the multidimensional Burgers equation.
Stretched vortex solutions of the Navier-Stokes equation are supposed to
play a fundamental role in the organization of turbulence. Stretching
has the effect to render a decaying vortex like the Lamb-Oseen vortex
into a stationary vortex (e.g. the Burgers vortex). Similar phenomena
are expected in the present case of the forced Burgers equation. 

As can be seen from the solvability conditions,
the crucial point is to obtain a solution of the nonlinear differential
equation for $a(t)$. The behaviour of $a(t)$ can be visualized
by introducing the potential
$V(a)=\frac{1}{3}a^3$: The variable $a(t)$ behaves as an overdamped 
particle moving in the potential $V(a)$ under the influence of the 
time dependent force $G(t)$. By introducing the variable $b(t)$ according to
\begin{equation}
a(t)=\frac{\dot b(t)}{b(t)}
\end{equation}
the equation of motion for $a(t)$ can be transformed 
into a linear equation of second order (Hill's equation)
\begin{equation}
\ddot b(t)= G(t) b(t) \qquad. 
\end{equation}
In terms of $b(t)$, the variables $\Lambda(t)$ and $\tau(t)$ are determined 
by
\begin{eqnarray}
\Lambda(t) &=& \frac{b(0)}{b(t)}
\nonumber \\
\tau(t)&=& \int_0^t dt' (\frac{b(0)}{b(t')})^2
\end{eqnarray}
In the following, we shall briefly consider some cases of functions $G(t)$. 

The first case is a stationary forcing, $G(t)=g$. 
If we take g to be a positive number, one obtains two solutions for a
\begin{equation}
a=\pm \sqrt{g} \qquad \Lambda = e^{\mp \sqrt{g}t} \qquad \tau = 
\frac{e^{\mp 2 \sqrt{g}t}-1}{\mp 2 \sqrt{g}} \qquad .
\end{equation}
Let us first consider the case of a positive value of $a=\sqrt{g}$. 
This solution is linearly stable. The contribution of a shock 
\begin{equation}
- A\Lambda(t) tanh \frac{A \Lambda(t)}{2\nu} r 
\end{equation}
decays to zero. A scale invariant solution also decays, since
\begin{equation}
\frac{\Lambda(t)}{\sqrt{\tau(t)}} W(\frac{\Lambda(t)}{\sqrt{\tau(t)}}r)
\end{equation}
and 
\begin{equation}
\lim_{t\rightarrow \infty} 
\frac{\Lambda(t)}{\sqrt{\tau(t)}}=2 \sqrt{g} \frac{e^{-2\sqrt{g}t}}
{1-e^{-2\sqrt{g}t}}=0
\end{equation}

The case of a negative value of $a=-\sqrt{g}$ is linearly unstable. 
Furthermore, shocklike contributions increase in time connected with a 
steepening of the shock. Scale invariant contributions in the long time limit 
become stationary since
\begin{equation}
\lim_{t\rightarrow \infty} 
\frac{\Lambda(t)}{\sqrt{\tau(t)}}=2 \sqrt{g} \frac{e^{2\sqrt{g}t}}
{e^{2\sqrt{g}t}-1}=\sqrt{2 \sqrt{g}}
\end{equation}  
such that the solution takes the form
\begin{equation}
v(r,t)=-\sqrt{g} r + \sqrt{2 \sqrt{g}} W(\sqrt{2 \sqrt{g}}r) \qquad .
\end{equation}

If we take $g=-\omega^2$ to be negative, we obtain an oscillating solution 
for $b(t)$ and a finite time singularity for $a(t)$:
\begin{equation}
b(t)= A ~cos ~\omega t \qquad a(t)= - \omega ~tan ~\omega t
\end{equation}
Furthermore, the quantity $\sigma(t)$ reads: 
\begin{equation}
\frac{1}{\sigma(t)}=\frac{\Lambda(t)}{\sqrt{\tau(t)}}=\sqrt{\frac{2\omega}
{sin 2\omega t}}
\end{equation}

We shall now consider the case where $G(t)$ is a stochastic white noise force.
The Fokker-Planck equation for the probability distribution
$f(a,t)=<\delta(a-a(t))>$ reads
\begin{equation}
\frac{\partial }{\partial t} f(a,t)=[
\frac{\partial }{\partial a} \frac{\partial V(a)}{\partial a} 
+Q \frac{\partial^2}{\partial a^2}]f(a,t)
\end{equation}
A stationary solution to this equation can be obtained: 
\begin{equation}
f_0(a)= N e^{-V(a)/Q}
\end{equation}
However, since the potential $V(a)$ is not bounded from below, this 
stationary solution does not define a probability distribution,
i.e. a stationary state does not exist. Due to the quadratic nonlinearity 
in the Langevin equation a kind of stochastic  
finite time singularity exists, where $a(t)$ tends to
$a(t) \rightarrow -\infty$. This can be seen from the fact that the variable $a$
moves like an overdamped particle in the unbounded potential under the
influence of a stochastic force.
As a consequence the variable $\Lambda(t)$ increases, too. 
 
Although there is a finite time singularity
it is quite interesting to investigate the structure of the probability 
distribution of the velocity field $v(r,t)$. To this end 
we introduce the distribution
$P(a,\Lambda,\tau;t)$, which obeys the Fokker-Planck equation
\begin{equation}
\frac{\partial }{\partial t} P= \lbrace
\frac{\partial }{\partial \Lambda} a \Lambda 
-\frac{\partial }{\partial \tau} \Lambda^2
+\frac{\partial }{\partial a} a^2
+\frac{\partial^2 }{\partial a^2} \rbrace P
\end{equation}
Using $P(a,\Lambda,\tau;t)$ the probability distribution of the
velocity field $v(r,t)$, $f(v,r,t)=<\delta(v-v(r,t)>$ takes the following form
\begin{eqnarray}
f(v,r,t) &=& \int d\Lambda d\tau da P(a,\Lambda,\tau,t)
\delta(v-a r -\Lambda w(\Lambda r,\tau))
\nonumber \\
&=& \frac{1}{r} \int d\tau d\Lambda 
P(\frac{v-\Lambda w(\Lambda r,\tau)}{r}, \Lambda,\tau,t)
\end{eqnarray}
The joint pdf for the velocity field at multiple scales $r_i$ takes the form
\begin{equation}\label{solv}
f(\lbrace v_i,r_i \rbrace,t)= \int d\Lambda d\tau da P(a,\Lambda,\tau,t)
\Pi_{i=1}^N \delta(v_i-a r_i -\Lambda w(\Lambda r_i,\tau)) \qquad .
\end{equation}
This formula demonstrates how the
spatial coherence of the stochastic force governs the
statistics of the multiple scales pdf of the velocity field.
  
Using standard methods one may easily derive an evolution equation for 
the pdf $f(v,r,t)$:
\begin{eqnarray}
\frac{\partial }{\partial t} f(v,r,t) &+&
v\frac{\partial }{\partial r} f(v,r,t)+
\frac{\partial }{\partial r} \int_{-\infty}^v dv'  f(v',r,t)
\nonumber \\
&=& -\nu \int dv' \int dr' \delta(r-r') v' \frac{\partial^2}{\partial r'^2}
f(v',r';v,r;t)
\nonumber \\
&+& Q(r,r) \frac{\partial^2}{\partial v^2} f(v,r,t) \qquad .
\end{eqnarray}
This equation is the first equation in a whole chain of equations
for the joint pdf of the velocity field at multiple scales,
\begin{eqnarray}\label{hier}
\frac{\partial }{\partial t} f(v_i,r_i,t) &+&
\sum_i v_i\frac{\partial }{\partial r_i} f(v_i,r_i,t)+
\sum_i \frac{\partial }{\partial r} \int_{-\infty}^{v_i} dv_i'  f(v_i',r_i,t)
\nonumber \\
&=& - \sum_i 
\nu \int dv' \int dr' \delta(r-r') v' \frac{\partial^2}{\partial r'^2}
f(v',r';v_i,r_i;t)
\nonumber \\
&+& \sum_{i,j}
Q(r_i,r_j) \frac{\partial^2}{\partial v_i \partial v_j} f(v_i,r_i,t)
\end{eqnarray}
Here, the matrix $Q(r_i,r_j)$ is given by 
\begin{equation}\label{small}
Q(r_i,r_j)=Q r_i r_j
\end{equation}
We present this hierarchy, since Polyakov \cite{Polyakov}
considered the same hierarchy for the case of a translational invariant
white noise forcing. Here, we could obtain a solution to this chain 
of equations by a direct calculation in terms of equation (\ref{solv}).

Let us now discuss the relationship of the above problem of forced
Burgers equation to the problem considered by Polyakov \cite{Polyakov}
of translational invariant white noise forcing: 
\begin{eqnarray}\label{Pol}
[\frac{\partial}{\partial t}+u(x,t) \frac{\partial}{ \partial x}]
u(x,t)
&=& \nu
\frac{\partial^2}{\partial x^2}u(x,t)+F(x,t) 
\nonumber \\
<F(x,t) F(x',t)> &=& q(x-x')\delta(t-t') \qquad .
\end{eqnarray}
Of interest are the statistical properties of the velocity increments
\begin{equation}
v_x(r,t)=u(x+r,t)-u(x,t) \qquad .
\end{equation}
 
The increment $v_x(r,t)$ obeys the evolution equation
\begin{eqnarray}\label{evol}
[\frac{\partial}{\partial t} &+& u(x,t) \frac{\partial}{ \partial x}]
v_x(r,t)+v_x(r,t) \frac{\partial }{\partial r} v_x(r,t)
\nonumber \\
&=& \nu[
\frac{\partial^2}{\partial r^2}v_x(r,t)-
\frac{\partial^2}{\partial r^2}v_x(r,t)|_{r=0}]
+f_x(r,t) 
\nonumber \\
f_x(r,t)&=& F(x+r,t)-F(x,t)
\nonumber \\
<f_x(r,t)f_x(r',t)> &=& q(r-r')-q(r)-q(r')+q(0)
\end{eqnarray}
The additional dissipative term arises since one has the condition
\begin{equation}
v_x(r=0,t)=0 \qquad .
\end{equation}
In a comoving coordinate system, i.e. considering the 
increments $v_{x(t)}(r,t)$, where the path $x(t)$ is defined by
\begin{equation}
\dot x(t)=u(x(t),t)
\end{equation}
we obtain the forced Burgers equation for the increment
\begin{eqnarray}\label{evol}
[\frac{\partial}{\partial t} &+& v(r,t) \frac{\partial}{ \partial r}]
v(x,t) \nonumber \\
&=& \nu[\frac{\partial^2}{\partial r^2}v(r,t)
-\frac{\partial^2}{\partial r^2}v(r,t)|_{r=0}]+G(r,t) \qquad ,
\end{eqnarray}
Here, $G(r,t)$ is given by
\begin{equation}
G(r,t)=F(x(t)+r,t)-F(x(t),t)
\end{equation}
 
Polyakov investigated the statistics of increments in the limit of
small values of $r$ and approximated the function $q(\xi)$ by
\begin{equation}\label{finalic}
q(\xi)=q(0)-\frac{Q}{2}\xi^2+..
\end{equation}
This relationship can only hold for $|\xi|<< \sqrt{q(0)}/Q$, since for large
values of $\xi$ $q(\xi) \rightarrow 0$. This approximation, however, is
equivalent to retain only the term linear in $r$ in the expression of the
fluctiuating force:
\begin{equation}\label{approxic}
G(r,t) = r G(t) + h.o.
\end{equation}

The function $G(t)$ is related to the stochastic force $F(x,t)$ via the
relationship
\begin{equation}
G(t) = \frac{\partial }{\partial x} F(x,t)|_{x=x(t)}
\end{equation}
In this approximation we recover our starting equation, eq. (\ref{evol}),
but now formulated for increments. That means, that 
the solutions are given by a similar expression compared to eq. (\ref{ansatz})
\begin{equation}
v(r,t)=a(t) r +\Lambda(t) \tilde w(\Lambda(t) r, \tau(t))
\end{equation}
provided we take $\tilde w(\xi,\tau)$ as increment solutions,
i.e. $\tilde w(\xi,\tau)= w(x+\xi,\tau)-w(\xi,\tau)$, where $w(\xi,\tau)$ 
is a solution of the homogeneous Burgers equation.

Let us now briefly consider the treatment of the velocity increment
statistics in the translationally invariant white noise forcing case, which 
has been performed by Polyakov
on the basis of a hierarchy similar to eq. (\ref{hier}). 
The only difference in this hierarchy is the form of the
matrix $Q(r_i,r_j)$. It is defined according to 
\begin{equation}
Q(r_i,r_j)=q(r_i)+q(r_j)-q(0)-q(r_i-r_j)
\end{equation}
Taking the approximation
$q(r)=q(0)-Q/2 r^2$ the same hierarchy arises since in this approximation
$Q(r_i,r_j)=Q r_i r_j$. Under this approximation the 
hierarchy of Polyakov 
is solved by the probability distributions given in eq. (\ref{solv}). 
However, as we have indicated above, the limit $t \rightarrow \infty$
of the solution does not exist. Therefore, the evaluation of the stationary
pdf's for the Polyakov problem has to go beyond the approximation of the 
force eq. (\ref{approxic}). This indicates that all calculations which have 
been performed so far for the translationally invariant white noise forced 
Burgers equation exclusively based on the approximation (\ref{finalic}) 
should lead to the above probability distributions, which, as we have 
indicated above, have no stationary time limit.

Let us now finally consider the multidimensional forced Burgers equation:
\begin{equation}
\frac{\partial }{\partial t} {\bf v}({\bf r},t)+
{\bf v}({\bf r},t)\cdot \nabla_r {\bf v}({\bf r},t)=
\nu \Delta {\bf v}({\bf r},t)+G(t) {\bf r}
\end{equation}
Here, $G(t)$ denotes a time dependent matrix. We look for solutions
of the form
\begin{equation}
{\bf v}=A(t) {\bf r} +{\bf U}({\bf r},t)
\end{equation}
The matrix $A(t)$ is determined in such a way as to cancel the force field.
This leads to the matrix equation
\begin{equation}
\dot A + A A = G(t) \qquad ,
\end{equation}
whereas the additional field ${\bf U}({\bf r},t)$ has to fulfill the
equation
\begin{eqnarray}
\frac{\partial }{\partial t} {\bf U}({\bf r},t)+A \cdot {\bf U}({\bf r},t)
+(A \cdot {\bf r})\cdot \nabla {\bf U}({\bf r},t)+ 
{\bf U}({\bf r},t)\cdot \nabla_r {\bf U}({\bf r},t)=
\nu \Delta {\bf U}({\bf r},t)
\end{eqnarray}
We consider the ansatz
\begin{equation}
{\bf U}({\bf r},t)={\bf W}(\Lambda(t) \cdot {\bf r},t)
\end{equation}
and assume the matrix $\Lambda(t)$ to obey the equation
\begin{equation}
\dot \Lambda^T + A^T \Lambda^T =0
\end{equation}
 This leads to the equation
\begin{equation}
\frac{\partial }{\partial t} {\bf W}({\bf \xi},t)+A \cdot 
{\bf W}({\bf \xi},t)
+ 
{\bf W}({\bf \xi},t)\cdot \Lambda^T \cdot \nabla_\xi {\bf W}({\bf \xi},t)=
\nu \nabla_\xi \Lambda \Lambda^T\cdot \nabla_\xi {\bf W}({\bf \xi},t)
\end{equation}
The ansatz
\begin{equation}
{\bf W}({\bf \xi},t)=\Phi(t){\bf w}({\bf \xi},t)
\end{equation}
where $\Phi(t)$ has to obey the equation
\begin{equation}
\dot \Phi(t) + A(t) \Phi(t) =0
\end{equation}
leads to 
\begin{equation}\label{genburg}
\frac{\partial }{\partial t} {\bf w}({\bf \xi},t)
+ 
{\bf w}({\bf \xi},t)\cdot \Phi^T
\Lambda^T \cdot \nabla_\xi {\bf w}({\bf \xi},t)=
\nu \nabla_\xi \Lambda \Lambda^T\cdot \nabla_\xi {\bf w}({\bf \xi},t)
\end{equation}
The matrix $\Phi(t)$, $\Lambda(t)$ obeys the equation
\begin{eqnarray}
\dot \Phi + A \Phi &=& 0
\nonumber \\
\dot \Lambda + \Lambda A &=& 0
\end{eqnarray}
We do not pursue the question further, what conditions have to be
imposed on the matrix $G(t)$ in order to allow for a complete reduction 
of equation (\ref{genburg}) to the Burgers equation (possibly in a
space with lower dimensions).
The simplest case is clearly the case of a diagonal matrix 
\begin{equation}
G(t)=g(t) E
\end{equation}
Then, all matrices $\Phi$, $\Phi^T$, $\Lambda$, $\Lambda^T$ are diagonal
matrices and, therefore, coincide. The matrices 
\begin{equation}
\Lambda \Lambda^T= \Phi^T \Lambda^T= \lambda^2(t) E
\end{equation}
are proportional to the unit matrix such that a transformation
\begin{equation}
\frac{d \tau(t)}{dt}=\lambda^2
\end{equation}
actually defines ${\bf w}({\bf \xi},\tau)$ as the solution of the 
multidimensional Burgers equation. We mention that a similar 
treatment has been performed for the case of stretched vortices in
hydrodynamic flows \cite{Lundgren}\\
    
Summarizing we have considered a forced Burgers equation with a forcing
linear in the spatial coordinate. We have shown that in the one dimensional
case this problem can be solved completely. We have discussed the 
relationship to the velocity increment statistics of the Burgers equation
with translationally invariant forcing. In the multidimensional case 
we have indicated the possibility of a similar treatment, which shows 
an analogy to the treatment of stretched vortices in hydrodynamic flows.

\end{document}